\documentclass{article}
\usepackage[utf8]{inputenc}
\usepackage[T1]{fontenc}
\usepackage[a4paper,left=15mm, right=15mm]{geometry}
\usepackage{graphicx} % Required for inserting images
\usepackage{physics}
\usepackage{siunitx}
\usepackage{hyperref}
\usepackage{amsmath}
\usepackage{amssymb}
\usepackage{csquotes}
\usepackage{todonotes}
\usepackage{booktabs}
\usepackage{authblk}
\usepackage{lineno}

\title{Interferometric measurement of nuclear resonant phase shift with a nanoscale Young double waveguide}
\author[1,2,3]{Leon M. Lohse\thanks{\texttt{leon.merten.lohse@desy.de}}}
\author[1,3]{Ankita Negi}
\author[2]{Markus Osterhoff}
\author[2]{Paul Meyer}
\author[4]{Sergey Yaroslavtsev}
\author[4]{Aleksandr I. Chumakov}
\author[3,1]{Lars Bocklage}
\author[6,7,1,3]{Ralf Röhlsberger}
\author[2]{Tim Salditt}
\date{5 June 2025}
\affil[1]{The Hamburg Centre for Ultrafast Imaging, Universität Hamburg, Hamburg, Germany}
\affil[2]{Georg-August-Universität Göttingen, Göttingen, Germany}
\affil[3]{Deutsches Elektronen-Synchrotron DESY, Hamburg, Germany}
\affil[4]{ESRF -- The European Synchrotron, Grenoble, France}
\affil[5]{Friedrich-Schiller-Universität Jena, Jena, Germany}
\affil[6]{Helmholtz-Institut Jena, Jena, Germany }
\affil[7]{GSI Helmholtzzentrum für Schwerionenforschung GmbH, Darmstadt, Germany}
\DeclareSIUnit\bar{bar}

\begin{document}

\maketitle

%\linenumbers
\paragraph{Abstract:}
The phase shift of an electromagnetic wave, imprinted by its interaction with atomic scatterers, is a central quantity in optics and photonics.
In particular, it encodes information about optical resonances and photon-matter interaction.
While being a routine task in the optical regime, interferometric measurements of phase shifts in the x-ray frequency regime are notoriously challenging due to the short wavelengths and associated stability requirements. As a result, the methods demonstrated to date are unsuitable for nanoscopic systems.
Here, we demonstrate a nanoscale interferometer, inspired by Young's double-slit experiment, to measure the dispersive phase shift due to the \SI{14.4}{\keV} nuclear resonance of the Mössbauer isotope $^{57}$Fe coupled to an x-ray waveguide. 
From the single-photon interference patterns, we precisely extract the phase shifts in the vicinity of the nuclear resonance resolved in photon energy by using Bayesian inference. 
We find that the combined information from phase shift and absorbance reveals microscopic coupling parameters, which are not accessible from the intensity data alone.
The demonstrated principle lays a basis for integrated x-ray interferometric sensors.

\section*{Introduction}

% the importance of phase
The phase of an electromagnetic wave is shifted upon scattering from an atom, a microscopic effect which adds up coherently and determines the macroscopic properties of wave propagation, as described by the index of refraction $n$. Near atomic transitions, resonant effects become particularly pronounced and 
a wealth of information about the photon-atom interactions and the involved atomic energy levels is encoded in the optical response.
Both for the resonant and non-resonant case, wave propagation phenomena are relevant for nearly all methods for the manipulation of light. Examples are plenty: from passive shaping with lenses and phase plates, guiding of light in fiber optics and waveguides, all the way to the precise active control, for example in liquid crystal displays, electro-optical modulators, and integrated photonic devices. 
These optical properties and functionalities crucially rely on precisely known and controllable phase shifts.
At the same time, when a wave propagates through an object with a spatially dependent index of refraction $n({\bf r})$, the local variations in the accumulated phase shifts reflect its internal structure, offering a contrast mechanism for phase-sensitive imaging.
Note that phase contrast can not only be significantly more prominent than attenuation contrast \cite{Hawkes_SH_2019}, for example in X-ray imaging of biological matter,
but also more dose-efficient \cite{Spiecker_O_2023}, because the underlying elastic scattering process does not deposit any energy in the specimen.

Guided modes offer a particular benefit to control light propagation and phase effects.  
In optical waveguides, photons can be coupled to individual quantum emitters such as resonant atoms, in a controllable manner. This is exploited in the emerging field of waveguide quantum electrodynamics in order to investigate quantum many-body physics and to develop integrated functional quantum technologies, see \cite{Tuerschmann_N_2019,Sheremet_RoMP_2023} for a review. The resulting quantum-optical systems are often non-linear \cite{Tuerschmann_N_2019}, and can become increasingly complex so that refractive index models are no longer valid. Yet the phase shift imposed by individual emitters remains a central quantity, which can even become a function of the number of photons \cite{Staunstrup_NC_2024}. Further, the phase shift can be manipulated, for example, in quantum phase switches with single atoms \cite{Tiecke_N_2014,Volz_NP_2014}, a building block that may pave the way to photon-photon quantum gates.
To observe any of this experimentally, one crucially relies on methods for precise and accurate measurements of phase shifts. 

% phase measurement in the hard x-ray regime
In the optical regime, instruments such as the Mach-Zehnder interferometer are readily available to measure phase shifts. They can be assembled on optical tables from  discrete components, but are also routinely realized in integrated photonic devices. In the x-ray regime on the other hand, the situation is quite different, because the short wavelengths require extreme stability and translation precision in the order of picometers. Bonse and Hart realized the first interferometer for hard x-rays \cite{Bonse_APL_1965}, a Mach-Zehnder-type setup, achieving the necessary stability by cutting the entire interferometer from one solid single crystal. This principle based on crystal reflections \cite{Colella__1996,Bowen__1996,Lider_P_2014} has enabled the design of   
interferometers that have been used to measure refractive indices and dispersion corrections to the atomic form factors close to absorption edges \cite{Bonse_ZfPAHan_1969,Templeton_ACSA_1980}.
In most cases, however, refractive indices in the hard x-ray regime are obtained indirectly from absorption spectra based on Kramers-Kronig relation \cite{Henke_ADaNDT_1993,Lengeler__1994},
a method which relies on precisely measured absorption spectra over a wide frequency range. 
Unfortunately, neither bulky crystal interferometers nor any of the other methods \cite{Lengeler__1994} are suitable to measure phase shifts at the nanometer scale.

This limitation is particularly regrettable, since x-ray photon-matter interaction within nanostructures exhibits many interesting phenomena \cite{Shi_PiQE_2024}.
These include x-ray generation in nanostructures \cite{Vassholz_SA_2021}, Purcell-enhanced x-ray scintillation, and enhanced interaction in waveguide-cavities \cite{Lentrodt__2024}.
Nanometer-sized x-ray waveguides have been realized for double-waveguide phase-contrast imaging \cite{Fuhse_PRL_2006}, curved x-ray waveguides \cite{Salditt_PRL_2015}, and integrated x-ray beam splitters \cite{HoffmannUrlaub_ACSA_2016}, representing first steps towards \enquote{x-ray optics on a chip}.
Recently, ensembles of Mössbauer nuclei coupled to an x-ray waveguide were experimentally investigated with synchrotron radiation, demonstrating a platform for waveguide quantum electrodynamics in the hard x-ray regime \cite{Andrejic_PRA_2024,Lohse__2024}.
Such Mössbauer isotopes constitute exceptionally clean quantum systems due to their extremely narrow resonances \cite{Roehlsberger__2021}, making them a particularly successful experimental platform for quantum-optics in the hard x-ray frequency regime \cite{Roehlsberger_S_2010,Heeg_N_2021,Khairulin_SR_2021,Bocklage_SA_2021,Shvyd’ko_N_2023,Velten_SA_2024}. 
Despite the tremendous importance of the phase shift in these examples, however, it has been elusive to direct measurement.

\begin{figure}[htb]
    \centering
    \includegraphics{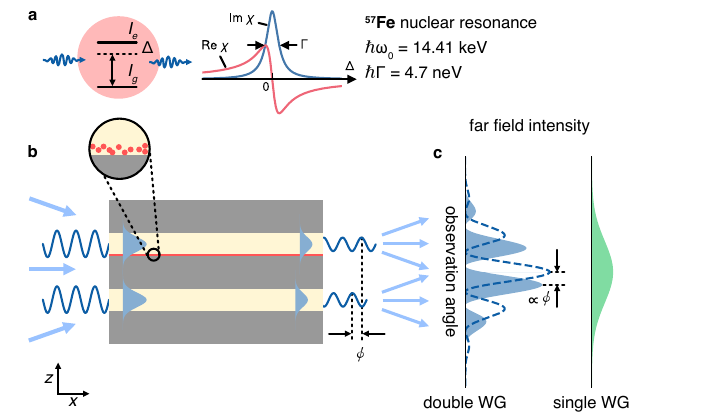}
    \caption{\textbf{Principle of the double-waveguide interferometer.} \textbf{a} A resonant atom (here, the nuclear resonance of $^{57}\mathrm{Fe}$ between states with nuclear spin $I_g = 1/2$ and $I_e = 3/2$) causes a phase shift depending on the detuning $\Delta = \omega - \omega_0$ of the light frequency $\omega$ to the resonance $\omega_0$. This is described by the real part of the complex susceptibility $\chi$. \textbf{b} An ensemble of resonant atoms (red) in a nanoscopic waveguide coherently shifts the phase of photons propagating through this \enquote{signal waveguide} (SWG) compared to a reference waveguide (RWG) without the atoms. \textbf{c} The phase shift results in a shift of the far-field double-slit interference pattern (filled blue curve) of the two exit waves compared to the unshifted pattern (dashed line). Blocking the reference waveguide results in a single-slit diffraction pattern (green filled curve) that is unaffected by the phase shift.}
    \label{fig:principle}
\end{figure}

% this work
In this work, we interferometrically measure the phase shift that an ultrathin layer of $^{57}\mathrm{Fe}$ Mössbauer nuclei coherently imprints onto photons propagating through a single-mode x-ray waveguide. To that end, we have devised a nanoscale double-waveguide interferometer, reminiscent of Young's celebrated double-slit experiment \cite{Young_PTotRSoL_1802}. It coherently superimposes the exit fields of the signal waveguide (SWG) and a reference waveguide (RWG) which is free of resonant nuclei, and directly converts the resonant phase shift in the SWG into a measurable lateral shift of the far-field interference pattern (see Fig.\ \ref{fig:principle}\textbf{b}).
This allows to extract the nuclear phase shift from just a small number of detected photons.
The isotope $^{57}\mathrm{Fe}$ has a low-lying nuclear resonance at $\hbar \omega_0 = \SI{14.41}{\keV}$ with a linewidth of only $\hbar \Gamma = \SI{4.7}{\nano\eV}$ (Fig.\ \ref{fig:principle}\textbf{a}). 
While several methods to measure the phase shift induced by macroscopic samples containing resonant nuclei have been proposed \cite{Sturhahn_PRB_2001} and demonstrated \cite{Sturhahn_ELE_2004,Callens_PRB_2005,Heeg_PRL_2015,Goerttler_PRL_2019,Wolff_PRR_2023,Yuan_NC_2025,Negi__2025}, they are all indirect methods that rely on computational reconstruction of the spectral phase shift from measured time-domain response functions, which requires a large number of detected photons. To that end, the sample and a resonant reference absorber are usually illuminated with short broad-band pulses and temporal interference patterns are recorded. 
By contrast, here we directly measure the spectral phase shift $\phi(\Delta)$ resolved in detuning from the resonance $\Delta = \omega - \omega_0$ using monochromatic narrowband illumination by coherently interfering individual photons. 
The measured phase shifts unambigously reveal the value of hyperfine broadening and the coupling strength between the nuclei and photons propagating in the waveguide, beyond the information that is accessible from absorption measurements alone.

\section*{Results}

%\paragraph{Sample:}
To implement the nano-interferometer, we have fabricated a pair of \SI{0.5}{\mm} long planar x-ray waveguides, stacked on top of each other, one of which (the signal waveguide) contains a \SI{0.6}{\nm}-thin film (about two atomic layers) of isotope-enriched $^{57}\mathrm{Fe}$ (see Fig. \ref{fig:principle}). The waveguides consist of a \SI{20}{\nm}-thick boron carbide ($\mathrm{B}_4\mathrm{C}$) guiding core surrounded by \SI{30}{\nm} of molybdenum cladding, resulting in the interferometer baseline of \SI{50}{\nm}. The design ensures that only a single guided mode of the electromagnetic field is transmitted by each waveguide. The waveguides are enclosed in \SI{1}{\mm} thick germanium wafers to block the tails of the illuminating focus \cite{Krueger_JoSR_2012}.
In addition, as control experiments, we have prepared a double-waveguide without the resonant nuclei, as well as a single waveguide containing resonant nuclei. The latter conceptually corresponds to blocking the reference \enquote{slit}.

\begin{figure}[htb]
    \centering
    \includegraphics{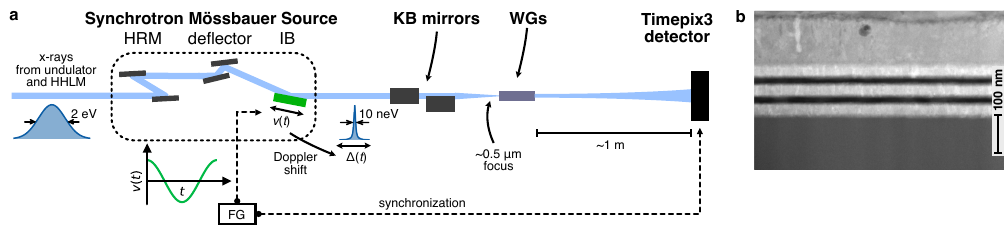}
    \caption{\textbf{Experimental scheme of the waveguide nano-interferometer.} \textbf{a} The broadband undulator radiation, pre-monochromatized to about \SI{2}{\eV} with a high heat load monochromator (HHLM), is further monochromatized to the nuclear linewidth by the Synchrotron Mössbauer source, then focussed by elliptical mirrors in Kirkpatrick-Baez (KB) geometry and front-coupled into the double-waveguide (WGs). The far-field pattern of the double-waveguide is observed by a time-resolving pixel detector (Timepix3). Internally, the SMS consists of a high-resolution monochromator (HRM), a deflector and the iron borate (IB) crystal. The IB is mounted on a velocity transducer, which periodically Doppler-shifts the photon energy, controlled by a function generator (FG). The detector is synchronized to the IB oscillation so that the photon time of arrival encodes the instantaneous frequency detuning. \textbf{b} Cross-sectional scanning transmission electron microscopy image of the double-waveguide.}
    \label{fig:setup}
\end{figure}

%\paragraph{Setup:}
The experiment requires a beam of hard x-rays at photon energy $\hbar \omega_0 = \SI{14.4}{\keV}$ which is both highly monochromatic, on the order of the natural linewidth of $^{57}\mathrm{Fe}$, and tightly focused, to couple efficiently into the front face of the two single-mode waveguides \cite{Fuhse_APL_2004}.
To this end, the experiments were performed at the nuclear resonance beamline ID18 \cite{Rueffer_HI_1996} of the European Synchrotron Radiation Facility (ESRF), utilizing the \enquote{Synchrotron Mössbauer Source} (SMS) \cite{Potapkin_JoSR_2012} to provide a monochromatic beam, which is subsequently focused to a spot of about \SI{0.5}{\um} in diameter by a pair of elliptical focusing mirrors in Kirkpatrick-Baez geometry \cite{Kupenko_HPR_2024}. A sketch of the experimental setup is shown in Fig.\ \ref{fig:setup}. The SMS employs an electronically forbidden but nuclear allowed Bragg reflection of an isotopically enriched iron borate ($^{57}\mathrm{FeBO}_3$) crystal to extract a bandwidth of a few neV with practically zero background \cite{Smirnov_PRA_2011}. To tune the photon energy over the target resonance, the iron borate is mounted on a Doppler velocity transducer, which is harmonically oscillating (the motion is harmonic for optimal stability), thus periodically shifting the energy via the Doppler effect. 
The photons transmitted by the waveguides are detected by a time-resolving pixel array detector based on Timepix3 \cite{Turecek_JoI_2016}. 
The detector is synchronized to the velocity transducer so that the photon time-of-arrival encodes the transducer oscillation phase and thereby the detuning of the photon energy. The detector is installed \SI{1}{\metre} downstream of the waveguides, where the \SI{55}{\um} pixel pitch provides an angular resolution of the interference pattern of about \ang{0.003}, with a total photon detection rate on the order of \SI{10}{\per\second}.

%\paragraph{Interference patterns:}
\begin{figure}[htb]
    \centering
    \includegraphics{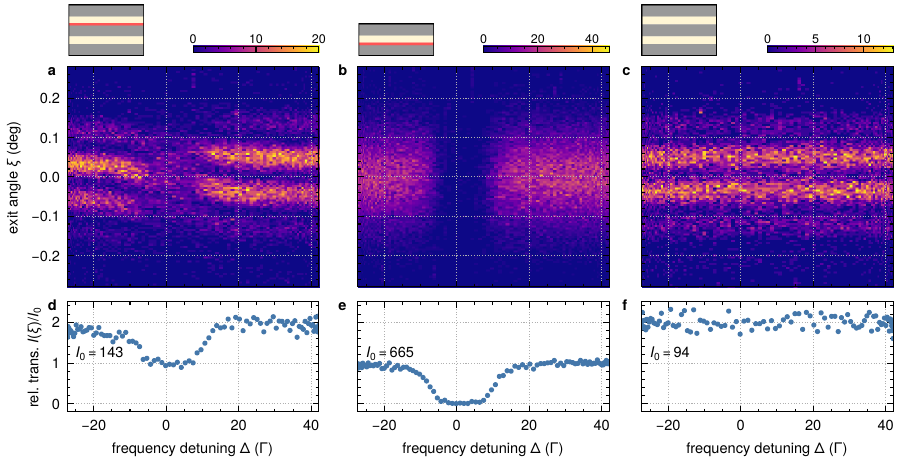}
    \caption{\textbf{Measured energy-resolved interference patterns and corresponding absorption spectra.}  The sketches illustrate the 3 different cases, within which the red line indicates the layer of Mössbauer nuclei. \textbf{a}-\textbf{c} Far-field interference pattern in terms of exit angle $\xi$ and detuning $\Delta = \omega - \omega_0$ for the three cases: \textbf{a} signal and reference waveguide, \textbf{b} signal waveguide only,
    \textbf{c} two reference waveguides (no Mössbauer nuclei).
    The color encodes the total number of detected photons per bin.  \textbf{d}-\textbf{f} Waveguide transmission coefficients (\enquote{Mössbauer absorption spectra}) for the 3 cases, respectively (column-wise sum of the interference patterns). $I_0$ indicates the total number of detected photons per interference pattern. }
    \label{fig:interference}
\end{figure}
Figure \ref{fig:interference} shows the measured far-field interference patterns as a function of photon energy detuning (see Methods for a detailed account of the data pre-processing) for the three interferometer
configurations: \textbf{a} \emph{with} resonant nuclei in the signal waveguide, \textbf{b} with nuclei in the signal waveguide
but reference waveguide blocked (no reference), and \textbf{c} both waveguides \emph{without} resonant nuclei.
Hence, \textbf{b} and \textbf{c} can be considered \enquote{control} experiments to the case of interest \textbf{a}. 
The lower panels show the corresponding absorption spectra, obtained by summing the interference pattern column-wise. The single waveguide \textbf{e} exhibits a broadened resonant absorption dip, which is reproduced by the interferometer \textbf{d} with a unit offset due to the reference waveguide.
Note that the interference patterns encode the phase difference between the fields at the exits of the two waveguides, including shifts accumulated within the waveguides as well as phase differences already present at the waveguide entrances. In particular, a slightly tilted wave front results in a frequency-independent shift of the interference pattern, a global offset. 
This effectively limited the maximum acquisition time and consequently the total number of detected photons in the interference patterns, because of gradual drifts of the wave front of the illuminating focus, likely due to the ambient temperature (see Methods).

The far-field interference pattern can be modeled by (see Methods)
\begin{equation}
\label{eq:interference}
    I(\xi) = I_0(\xi) \left[ 1 + T + V_0 \sqrt{T} \cos(\phi + k b \xi) \right],
\end{equation}
where $I_0(\xi)$ is the far-field diffraction pattern of a single waveguide mode, $\xi$ is the observation angle, $b$ the baseline (waveguide separation in $z$), $\phi$ the phase shift between the two waveguides, $k = \omega_0/c$ the on-resonance wavenumber, and $T$ the intensity transmission coefficient of the SWG. The parameter $V_0 \le 1$ accounts for a global loss of fringe visibility.
On resonance, the nuclei fully absorb the photons ($T = 0$) which results in 
\SI{0}{\percent} and \SI{50}{\percent} transmission in the single and double waveguide, respectively. Further away from resonance, the interference pattern is clearly visible and continuously shifts with varying detuning.

%\paragraph{Phase shifts:}
From the measured interference pattern, we have extracted the phase shifts $\phi(\Delta)$ induced by the nuclear resonance, using Bayesian inference \cite{Gelman__2004} taking the Poissonian probability distribution of the data into account.
To that end, we have computed the full \emph{posterior probability density} of the phase shift $p(\phi \mid \mathrm{data})$, conditional on the respective observed interference pattern, using a Markov Chain Monte Carlo ensemble sampler \cite{ForemanMackey_PotASotP_2013}, independently for each frequency detuning $\Delta$. This approach gives us not only the most likely value of the phase shift but its entire probability distribution and hence a robust estimate of the uncertainty.

\begin{figure}[htb]
    \centering
    \includegraphics{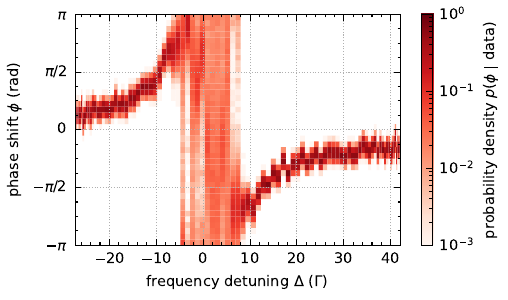}
    \caption{\textbf{Extracted phase shifts.} Phase shift $\phi$ extracted from the data shown in Fig.\ \ref{fig:interference}\textbf{a} by the Bayesian phase retrieval procedure. The color encodes the posterior probability density for $\phi(\Delta)$ given the measured interference patterns at a certain detuning $\Delta$. Due to the strong absorption, the phase is undetermined in a frequency band of $\pm 6\, \Gamma$ around the resonance frequency. Outside of that band, the phase is well determined.}
    \label{fig:phase}
\end{figure}
Figure \ref{fig:phase} shows the resulting probability densities for the phase shift. For larger detunings, the phase shift is precisely localized (standard deviation of \SI{0.13}{rad}) and exhibits the expected resonance. On the other hand, within $\pm 6\, \Gamma$ of the resonance, the visibility is practically zero (compare Fig. \ref{fig:interference}\textbf{a}) so that the interference pattern contains no information on the phase shift. As can be seen by the values of $I_0$ in Fig.\ \ref{fig:interference}\textbf{d}, the phases were extracted from interference patterns with less than 300 photons each.

%\paragraph{Theoretical model:}
\subsection*{Connecting the phase shifts to photon-matter interaction}
The phase shift contains information about the photon-nuclei coupling and helps to disambiguate parameters that lead to similar absorption spectra.
Based on a recently developed theoretical model \cite{Andrejic_PRA_2024}, the harmonic components of the field which is propagating (along $x$) through the waveguide while coherently interacting with the nuclei can be written as
\begin{equation}
\label{eq:field}
    E(\omega, x) = E_0(\omega, x) \exp\left[i k \zeta_m \chi(\omega) x/2\right],
\end{equation}
where $E_0$ is the reference field without nuclei and $\zeta_m$ quantifies the coupling between the single guided mode and the nuclear layer (see Methods). 
The bare nuclear resonance corresponds to the magnetic susceptibility
\begin{equation}
\label{eq:susceptibility}
    \chi_0(\omega) = - \rho_\mathrm{n} \frac{\sigma_\mathrm{rad}}{k} \frac{\Gamma / 2}{\omega - \omega_0 + i \Gamma/2}
\end{equation}
($\sigma_\mathrm{rad}$: radiative cross section, $k = \omega_0 /c$: wavenumber, $\rho_n$: number density of resonant nuclei) as shown in Fig.\ \ref{fig:principle}\textbf{a}. 
The nuclear levels are generally split due to electric and magnetic hyperfine interactions, leading to a modified susceptibility $\chi$. Since the thin iron layer, which is only about two atomic layers thick, does not develop long range magnetic order at room temperature, we do not expect magnetic hyperfine splitting here. Yet, even small differences of the environment and electric field gradients of the nuclei in the thin iron film lead to inhomogeneous line broadening and possibly a quadrupole splitting of the resonance line. We assume a Gaussian broadened susceptibility $\chi$ governed by two effective hyperfine parameters---net central shift and line broadening (see methods). 
\begin{figure}[htb]
    \centering
    \includegraphics{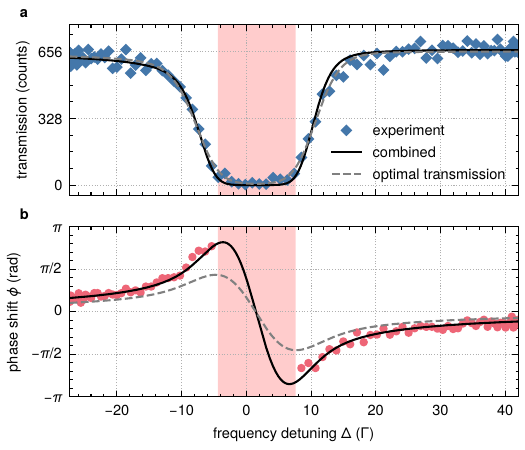}
    \caption{\textbf{Absorption spectra and phase shifts.} \textbf{a} Measured intensity transmission as a function of detuning $\Delta = \omega - \omega_0$ together with two model curves (see main text). \textbf{b} Extracted phase shifts (first moment of the posterior probability density) together with the corresponding model curves from \textbf{a}. The red area indicates the $\pm 6\ \Gamma$ band of nearly full absorption, for which the data does not contain any phase information.}
    \label{fig:fits}
\end{figure}

Figure \ref{fig:fits}\textbf{a} shows the measured absorption spectrum and two theory curves based on \eqref{eq:field}, for which the hyperfine parameters and coupling strength $\abs{\zeta_m}$ were found with maximum likelihood estimations with respect to either the absorption spectrum alone (dashed line), or the absorption spectrum combined with the phase shifts (solid line). More precisely, the former shows the model for which the measured absorption spectrum is most likely, with no regard for the phase shift. For the latter, on the other hand, the two hyperfine parameters were optimized with respect to the absorption spectrum and the coupling strength with respect to the phase shift. In both cases, only the modulus $\abs{\zeta_m}$ of the complex coupling coefficient $\zeta_m = \abs{\zeta_m} \exp\bqty{i \arg \zeta_m}$ was varied and the complex argument was fixed at the value $\arg \zeta_m^\dagger = \num{0.03}$, which was obtained from simulations using the design parameters of the layer structure.
\begin{table}[tbh]
    \centering
    \caption{\textbf{Hyperfine parameters and coupling strength extracted from the data.} The parameters ($\abs{\zeta_m}$: coupling strength, $w_\mathrm{HF}$: line broadening (FWHM), and $\delta_\mathrm{HF}$: central shift with respect to $\alpha$-iron) were found numerically via maximum likelihood estimation from the absorption spectrum alone, or the absorption spectrum combined with the phase shifts (see main text).}
    \label{tab:hyperfine}
    \begin{tabular}{l c c c}
    \toprule
       model & $\abs{\zeta_m}$ & $w_\mathrm{HF}$ &$\delta_\mathrm{HF}$ \\
       \midrule
       absorption & \num{4.5e-3} & $11\,\Gamma$ & $\num{1.6}\,\Gamma$ \\ 
       combined  & \num{7.1e-3} & $9\,\Gamma$ & $\num{1.6}\,\Gamma$ \\
       \bottomrule
    \end{tabular}
\end{table}

The parameters extracted in this way are listed in Tab.\ \ref{tab:hyperfine}. 
The coupling strength of the combined model closely matches the design value from theory calculations of $\abs{\zeta_m^\dagger} = \num{7.1e-3}$ (see methods), while the absorption model corresponds to substantially smaller coupling of \num{4.5e-3}.
Both models describe the absorption spectrum (Fig.\ \ref{fig:fits}\textbf{a}) well, which is the result of line broadening and the coupling strength being anticorrelated with respect to the absorption spectra.
In the phase shifts (Fig.\ \ref{fig:fits}\textbf{b}), however, we can clearly disambiguate the two models. It is evident that the first model does not reproduce the phase shifts at all, while the second model does. Thus, only the combination of the two observables allows us to uniquely extract the model parameters and in particular quantify the coupling parameter. 

The higher sensitivity and information content of the phase compared to the absorption signal can be attributed to the fact that  the phase shift asymptotically decays like $\phi \propto \Delta^{-1}$, whereas the absorbance decays like $\mu \propto \Delta^{-2}$. Consequently, even noisy measurements of the phase shift reveal quantitative information over a frequency range much wider than the width of the absorption line.

\section*{Discussion}
In summary, we have directly measured the resonant phase shift imprinted onto photons that propagate through an x-ray waveguide, while coherently interacting with an ensemble of nuclear resonances, using a nano-interferometer operated in the single photon regime. 
The method is based on two-path interference of individual photons and has allowed us to accurately extract the phase shifts from only on the order of 100 detected photons (for each energy detuning value) by employing Bayesian inference.
Using the extracted phase shift, we were able to quantify the coupling strength between waveguide mode and the nuclei, which was not possible from the absorption spectrum alone, since line broadening and optical thickness lead to ambiguous signatures (see Fig.\ \ref{fig:fits}).
The observed phase-shift and absorption spectra are consistently described by a recently derived theoretical model \cite{Andrejic_PRA_2024,Lohse_OE_2024}.

Our results demonstrate the interferometric measurement of phase shifts in the x-ray regime on the nanometer scale, using a few individual photons. The technique is not restricted to nuclear resonances but can analogously be applied to electronic resonances.
This paves the way for a new generation of \enquote{x-ray photonics on a chip}, combining curved waveguides \cite{Salditt_PRL_2015}, beamsplitters \cite{HoffmannUrlaub_ACSA_2016}, and x-ray sources \cite{Vassholz_SA_2021} with integrated interferometric sensors. 

Having accurately measured the phase shift, one can envision its control. A thicker layer of resonant iron nuclei will develop long range magnetic order so that the degenerate resonance splits into a sextet. The phase shift for a fixed photon energy could then be tuned by controlling the magnetic hyperfine spectrum, either by applying a static magnetic field or transiently with RF pulses as demonstrated in Ref.\ \cite{Bocklage_SA_2021}. This would enable active control of the phase for narrowband pulses in a nanophotonic device in the hard x-ray frequency range.

\clearpage
\newpage

\section*{Methods}

\subsection*{Waveguide design}
The waveguides were designed using the xwglib software from Ref.\ \cite{Lohse_OE_2024}. Materials, layer thicknesses, and length was chosen to effectively absorb everything but a single guided mode of the electromagnetic field. The iron layer was dimensioned to produce a resonant phase shift in the order of $\pm \pi$.

\subsection*{Waveguide fabrication}
The waveguide samples were fabricated via magnetron sputter deposition, using an argon plasma at a pressure of \SI{5e-3}{\milli\bar} at a target-to-substrate distance of about \SI{7}{\cm} with sputtering guns from the A300-XP series (AJA Inter-
national, Massachusetts, USA). The residual base pressure prior to deposition was typically less than \SI{4e-7}{\milli\bar}.
Molybdenum and the $^{57}\mathrm{Fe}$-isotope-enriched iron (\SI{97}{\percent}) were deposited using targets with diameter \SI{3.81}{\cm} at \SI{12}{\watt} DC, and $\mathrm{B}_4\mathrm{C}$ using a target with diameter \SI{5.08}{\cm} at \SI{53}{\watt}  RF power. The deposition rates were calibrated based on x-ray reflectivity measurements.
The waveguides were deposited onto \SI{1}{\mm} thick polished germanium wafers.
To block the over-illuminating part of the x-ray focus. A second \SI{1}{\mm}-thick germanium wafer was bonded on top of the layer structure, using soft alloy bonding as detailed in Ref. \cite{Krueger_JoSR_2012}. 

High-angle annular dark-field (HAADF-STEM) imaging was performed in an FEI Titan 80-300 E-TEM using an acceleration voltage of \SI{300}{\kilo\volt} and a convergence semi-angle of \SI{10}{mrad} yielding a probe size of approximately \SI{0.14}{\nano\metre}.

\subsection*{Event-based detection}
To detect the photons in the far-field, we used an AdvaPIX TPX3 (ADVACAM s.r.o., Czech Republic) with a \SI{500}{\um}-thick silicon sensor (\SI{70}{\percent} efficiency at \SI{14.4}{\keV} photon energy). We operated this Timepix3-based \cite{Turecek_JoI_2016,Poikela_JoI_2014} detector in event-based mode, so that it produces a list of detection events with pixel coordinates, time-of-arrival (ToA), and time-over-threshold data.
The velocity transducer of the SMS was controlled by a DFG-1200 digital function generator (Wissenschaftliche Elektronik GmbH, Germany) set to sinusoidal mode. We fed the TTL output of the DFG through a custom-built signal converter into the external synchronization interface of the AdvaPIX. In this way, we set the detector to reset its ToA counter in each oscillation period of the velocity transducer. The temporal resolution of the synchronization was limited by the detector clock frequency of \SI{40}{\MHz}---significantly exceeding the resolution required for the \SI{10}{\Hz} oscillation of the velocity transducer. As a result, the ToA data of the detected photons encodes the velocity of the transducer and thereby the energy detuning.

The total number of detected photons is limited by the fact that the interference pattern is highly sensitive to small relative changes in the wave front (phase) of the illuminating focus. It could only be kept stable for about \SI{1}{\hour} because the beamline optics had to be periodically realigned due to temperature drifts in the building, which slightly altered the wave front in the focus and shifted the interference patterns. We remark that this instability has been substantially reduced shortly after our experiment, as the beamline has been moved into temperature-controlled hutches.

\subsection*{Photon energy calibration}
The photon energy detuning was calibrated based on the absorption spectrum of a natural $\alpha$-iron reference absorber foil used for synchrotron and conventional Mössbauer spectroscopy. Zero detuning ($\Delta = 0$) corresponds to the center of the $\alpha$-iron absorption spectrum.

\subsection*{Data preprocessing}
Several preprocessing steps were applied to the data to generate the interference patterns.
\begin{enumerate}
    \item Apply a photon-energy filter to reduce spurious events due to environmental and cosmic radiation as well as charge-sharing events. For this, events outside a time-over-threshold range were discarded.
    \item Select data from a time window during which the beam was most stable by visual inspection.
    \item Select a horizontal range of the data that is as homogeneous as possible by visual inspection. This was necessary because the illuminating wave field had a visibly tilted phase gradient.
    \item Project the detection position onto the vertical axis.
    \item Convert ToA data to energy detuning based on the calibration and vertical pixel position $z$ to raw observation angle, i.e.\ $\xi_\mathrm{raw} = \arctan(z/D)$ where $D = \SI{1.18}{\metre}$ is the distance between waveguide and detector and $z$ the vertical position on the detector.
\end{enumerate}

\subsection*{Interference pattern}
Two (partially) coherent wave fields $E_1(x_0, z) = A_0(z+b/2)$ and $E_2(x_0, z) = A_0(z-b/2) \sqrt{T} e^{i \phi}$ with wavelength $\lambda$ centered at $z = \pm b/2$, where $A_0(z)$ describes a complex amplitude with finite width $w$, produce a far-field Fraunhofer diffraction pattern that can be described by equation \eqref{eq:interference} (see Ref.\ \cite{Saleh__2019}, for example).
The fringe visibility parameter $V_0 \in [0,1]$ accounts for the partial coherence \cite{Born__2019} and $I_0(\xi)$ is the Fraunhofer diffraction pattern of $A_0(z)$.
Setting $T_\mathrm{eff} = 1 + T$ and $V_\mathrm{eff} = \sqrt{T} V_0$ yields 
\begin{equation}
\label{eq:bayesmodel}
    I(\xi; \vec \theta) = I_0(\xi) T_\mathrm{eff} \left[ 1 + V_\mathrm{eff} \cos(\phi + k b \xi) \right],
\end{equation}
with model parameters $\vec \theta = (\phi, T_\mathrm{eff}, V_\mathrm{eff})$ and global (constant) parameters: wave number $k=2 \pi / \lambda$, mode width $w$, waveguide separation $b$, central angle $\xi_0$, total intensity $I_\mathrm{tot}$.
We approximate the exit field of the waveguide to be Gaussian, so that its far-field intensity distribution is
\begin{equation}
    I_0(\xi) = \frac{I_\mathrm{tot}}{\sigma \sqrt{\pi}} \exp[- (\xi - \xi_0)^2 / \sigma^2]
\end{equation}
where $\sigma = \num{2.355}/(kw)$.

\subsection*{Statistical model}
We employ a statistical model to implement phase retrieval from the sparse interference patterns.
The model parameters $\vec \theta$ will be extracted for each detuning $\Delta$ independently.
The probability to measure $y_j$ counts at angle $\xi_j$ conditional on certain values of $\vec \theta$ is
$p(y_j \mid \vec \theta) = \left.\operatorname{Pr}(Y_j = y_j \mid \lambda)\right|_{\lambda = I(\xi_j; \vec \theta)}$,
where
\begin{equation}
   \operatorname{Pr}(Y_j = y_j \mid \lambda) = \frac{\lambda^y_j e^{-\lambda}}{y_j!} 
\end{equation}
is the Poisson distribution. Since measurements for different angles are independent, we have $p(\vec y \mid \vec \theta) = \prod_j p(y_j \mid \vec \theta)$.
Bayes theorem implies for the \emph{posterior} distribution
\begin{equation}
\label{eq:bayes}
    p(\vec \theta \mid \vec y) \propto p(\vec \theta) \cdot p(\vec y \mid \vec \theta),
\end{equation}
where the proportionality constant only depends on the data $\vec y$ and is independent of the parameters $\vec \theta$ \cite{Gelman__2004}.
We use the \enquote{emcee} Markov Chain Monte Carlo (MCMC) ensemble sampler \cite{ForemanMackey_PotASotP_2013} to compute the posterior as detailed below.

The MCMC sampler is based on the logarithm of the right hand side of \eqref{eq:bayes}, which can be computed as follows.
The log-likelihood for a given angle $\xi$ is
\begin{equation}
\begin{split}
    \ell_1(\vec \theta; y) &\equiv \log p(y \mid \vec \theta )  \\
    &=  - I(\xi; \vec \theta) + y \log I(\xi; \vec \theta) - \log(y!).
\end{split}
\end{equation}
The total log-likelihood is therefore
\begin{equation}
    \ell(\vec \theta; \vec y) = \sum_j \left[ - I(\xi_j; \vec \theta) + y_j \log I(\xi_j; \vec \theta)\right] + \text{const.}
\end{equation}
We assume uniform prior distributions as follows:
$\phi \sim U(-\pi, \pi)$, $V_\mathrm{eff} \sim U(0, 1)$, and $T_\mathrm{eff}\sim U(\num{0.3}, \num{1.3})$. While the real part of the exponent in \eqref{eq:field} can take any value, the interferometer signal is $2 \pi$ periodic and hence, the phase is only uniquely retrievable within an interval of length $2 \pi$.

\subsection*{Phase retrieval}
Phase retrieval proceeds as follows.
\begin{enumerate}
    \item Estimate $\xi_0$, $w$ and $I_\mathrm{tot}$ from the interference patterns on resonance with $V_\mathrm{eff} = 0$ and $T_\mathrm{eff} = 1$.
    \item Estimate $b$ from a narrow range of off-resonance interference patterns, using the parameters from step 1.
    \item Compute a sample ensemble of $\vec \theta$ given the data $\vec y$ with \enquote{emcee}, using the global parameters of step 1 and 2.
    \item Estimate the probability density of the phase shift parameter $\phi$ from a marginal histogram of the ensemble.
    \item Shift the phase shift globally such that $\operatorname{E}[\phi(+\Delta)] = -\operatorname{E}[\phi(-\Delta)]$ for $\Delta = 20\ \Gamma$.
\end{enumerate}

\subsection*{Coupling constant}
The coupling constant $\zeta_m$ in \eqref{eq:field} is defined as
\begin{equation}
    \zeta_m = \frac{[u_m(z_0)]^2}{\nu_m} d,
\end{equation}
where $\nu_m$ is the (complex) effective refractive index of the mode, $d$ the thickness of the iron layer, $z_0$ the position of the iron layer, and $u_m(z)$ is the transversal profile of the guided mode, which is bi-normalized to $\int [u_m(z)]^2 / \epsilon(z) \dd z = 1$ ($\epsilon(z)$ being the relative electric permeability at height $z$ in the layer system) \cite{Lohse_OE_2024}. It can be computed numerically, for example with the xwglib library (see Ref.\ \cite{Lohse_OE_2024}).

\subsection*{Hyperfine splitting}
The \SI{0.6}{\nm}-thin iron layer is amorphous and too thin to develop long range-magnetic order. Thus, it does not exhibit magnetic hyperfine splitting. However, the individual nuclei can generally experience different isomeric shifts and possibly quadrupole splittings, due to their inhomogeneous environments in the film, which is only about two atomic layers thick. We account for this by introducing an effective Gaussian line broadening and a net central shift,
\begin{equation}
    \chi(\omega; \delta_\mathrm{HF}, \sigma_\mathrm{HF}) = f(\omega; \sigma_\mathrm{HF}) * \chi_0(\omega - \delta_\mathrm{HF})
\end{equation}
with
\begin{equation}
f(\omega; \sigma_\mathrm{HF}) = \frac{1}{\sigma_\mathrm{HF} \sqrt{2 \pi}} \exp\left\{-\frac{1}{2} \left(\frac{\omega}{\sigma_\mathrm{HF} } \right)^2\right\}.
\end{equation}

\section*{Acknowledgements}
We thank Mike Kanbach for preparing the waveguide assemblies, Matthias Hahn and Tobias Meyer for providing and supporting the electron microscope, Ilya Kupenko for supporting the nanoscope at ID18, and Dieter Lott and Helmholtz-Zentrum Hereon for providing and supporting the reflectometer to calibrate the thin film samples. 
We acknowledge the European Synchrotron Radiation Facility (ESRF) for provision of synchrotron radiation facilities at ID18 under proposal number MI-1454.
We acknowledge partial funding by Max Planck School of Photonics and Deutsche Forschungsgemeinschaft (DFG) (432680300 SFB 1456/C03; 390715994 EXC 2056). LML, PM, RR, and TS are part of the Max Planck School of Photonics.

\bibliographystyle{unsrt}
\bibliography{interferometer}

\begin{thebibliography}{10}

\bibitem{Hawkes_SH_2019}
Peter~W. Hawkes and John C.~H. Spence, editors.
\newblock {\em Springer Handbook of Microscopy}.
\newblock Springer International Publishing, 2019.

\bibitem{Spiecker_O_2023}
Rebecca Spiecker, Pauline Pfeiffer, Adyasha Biswal, Mykola Shcherbinin, Martin
  Spiecker, Holger Hessdorfer, Mathias Hurst, Yaroslav Zharov, Valerio
  Bellucci, Tom{\'{a}}{\v{s}} Farag{\'{o}}, Marcus Zuber, Annette Herz,
  Angelica Cecilia, Mateusz Czyzycki, Carlos Sato~Baraldi Dias, Dmitri Novikov,
  Lars Krogmann, Elias Hamann, Thomas van~de Kamp, and Tilo Baumbach.
\newblock Dose-efficient in vivo x-ray phase contrast imaging at micrometer
  resolution by bragg magnifiers.
\newblock {\em Optica}, 10(12):1633, December 2023.

\bibitem{Tuerschmann_N_2019}
Pierre T{\"{u}}rschmann, Hanna~Le Jeannic, Signe~F. Simonsen, Harald~R. Haakh,
  Stephan G{\"{o}}tzinger, Vahid Sandoghdar, Peter Lodahl, and Nir Rotenberg.
\newblock Coherent nonlinear optics of quantum emitters in nanophotonic
  waveguides.
\newblock {\em Nanophotonics}, 8(10):1641--1657, August 2019.

\bibitem{Sheremet_RoMP_2023}
Alexandra~S. Sheremet, Mihail~I. Petrov, Ivan~V. Iorsh, Alexander~V.
  Poshakinskiy, and Alexander~N. Poddubny.
\newblock Waveguide quantum electrodynamics: Collective radiance and
  photon-photon correlations.
\newblock {\em Reviews of Modern Physics}, 95(1):015002, March 2023.

\bibitem{Staunstrup_NC_2024}
Mathias J.~R. Staunstrup, Alexey Tiranov, Ying Wang, Sven Scholz, Andreas~D.
  Wieck, Arne Ludwig, Leonardo Midolo, Nir Rotenberg, Peter Lodahl, and Hanna
  Le~Jeannic.
\newblock Direct observation of a few-photon phase shift induced by a single
  quantum emitter in a waveguide.
\newblock {\em Nature Communications}, 15(1), August 2024.

\bibitem{Tiecke_N_2014}
T.~G. Tiecke, J.~D. Thompson, N.~P. de~Leon, L.~R. Liu, V.~Vuleti{\'{c}}, and
  M.~D. Lukin.
\newblock Nanophotonic quantum phase switch with a single atom.
\newblock {\em Nature}, 508(7495):241--244, April 2014.

\bibitem{Volz_NP_2014}
J{\"{u}}rgen Volz, Michael Scheucher, Christian Junge, and Arno Rauschenbeutel.
\newblock Nonlinear $\pi$ phase shift for single fibre-guided photons
  interacting with a single resonator-enhanced atom.
\newblock {\em Nature Photonics}, 8(12):965--970, November 2014.

\bibitem{Bonse_APL_1965}
U.~Bonse and M.~Hart.
\newblock An {X}-ray interferometer.
\newblock {\em Applied Physics Letters}, 6(8):155--156, April 1965.

\bibitem{Colella__1996}
R.~Colella.
\newblock X-ray and neutron interferometry: Basic principles and applications.
\newblock In Andr{\'{e}} Authier, Stefano Lagomarsino, and Brian~K. Tanner,
  editors, {\em X-Ray and Neutron Dynamical Diffraction: Theory and
  Applications}, volume 357 of {\em NATO Science Series B}, pages 369--380.
  Springer US, 1996.

\bibitem{Bowen__1996}
D.~K. Bowen.
\newblock Applications of x-ray interferometry.
\newblock In Andr{\'{e}} Authier, Stefano Lagomarsino, and Brian~K. Tanner,
  editors, {\em X-Ray and Neutron Dynamical Diffraction: Theory and
  Applications}, volume 357 of {\em NATO Science Series B}, pages 381--410.
  Springer US, 1996.

\bibitem{Lider_P_2014}
V.~V. Lider.
\newblock X-ray crystal interferometers.
\newblock {\em Physics-Uspekhi}, 57(11):1099--1117, November 2014.

\bibitem{Bonse_ZfPAHan_1969}
U.~Bonse and H.~Hellk{\"{o}}tter.
\newblock Interferometrische messung des brechungsindex f{\"{u}}r
  r{\"{o}}ntgenstrahlen.
\newblock {\em Zeitschrift f{\"{u}}r Physik A Hadrons and nuclei},
  223(4):345--352, August 1969.

\bibitem{Templeton_ACSA_1980}
D.~H. Templeton, L.~K. Templeton, J.~C. Phillips, and K.~O. Hodgson.
\newblock Anomalous scattering of x-rays by cesium and cobalt measured with
  synchrotron radiation.
\newblock {\em Acta Crystallographica Section A}, 36(3):436--442, May 1980.

\bibitem{Henke_ADaNDT_1993}
B.~L. Henke, E.~M. Gullikson, and J.~C. Davis.
\newblock X-ray interactions: Photoabsorption, scattering, transmission, and
  reflection at e = 50-30,000 ev, z = 1-92.
\newblock {\em Atomic Data and Nuclear Data Tables}, 54(2):181--342, July 1993.

\bibitem{Lengeler__1994}
B.~Lengeler.
\newblock Experimental determination of the dispersion correction {f'(E)} to
  the atomic scattering factor.
\newblock In G.~Materlik, C.~J. Sparks, and K.~Fischer, editors, {\em Resonant
  Anomalous X-ray Scattering: Theory and Applications}, pages 35--60.
  North-Holland, Amsterdam [u.a.], 1994.

\bibitem{Shi_PiQE_2024}
Xihang Shi, Wen~Wei Lee, Aviv Karnieli, Leon~Merten Lohse, Alexey Gorlach, Lee
  Wei~Wesley Wong, Tim Salditt, Shanhui Fan, Ido Kaminer, and Liang~Jie Wong.
\newblock Quantum nanophotonics with energetic particles: X-rays and free
  electrons.
\newblock {\em Progress in Quantum Electronics}, 2024.
\newblock In press.

\bibitem{Vassholz_SA_2021}
Malte Vassholz and Tim Salditt.
\newblock Observation of electron-induced characteristic x-ray and
  bremsstrahlung radiation from a waveguide cavity.
\newblock {\em Science Advances}, 7(4), January 2021.

\bibitem{Lentrodt__2024}
Dominik Lentrodt, Christoph~H. Keitel, and J{\"{o}}rg Evers.
\newblock Towards nonlinear optics with m{\"{o}}ssbauer nuclei using x-ray
  cavities, 2024.

\bibitem{Fuhse_PRL_2006}
C.~Fuhse, C.~Ollinger, and T.~Salditt.
\newblock Waveguide-based off-axis holography with hard x rays.
\newblock {\em Physical Review Letters}, 97(25):254801, December 2006.

\bibitem{Salditt_PRL_2015}
T.~Salditt, S.~Hoffmann, M.~Vassholz, J.~Haber, M.~Osterhoff, and J.~Hilhorst.
\newblock X-ray optics on a chip: Guiding x rays in curved channels.
\newblock {\em Physical Review Letters}, 115(20):203902, November 2015.

\bibitem{HoffmannUrlaub_ACSA_2016}
Sarah Hoffmann-Urlaub and Tim Salditt.
\newblock Miniaturized beamsplitters realized by x-ray waveguides.
\newblock {\em Acta Crystallographica Section A}, 72(5):515--522, August 2016.

\bibitem{Andrejic_PRA_2024}
Petar Andreji{\'{c}}, Leon~Merten Lohse, and Adriana P{\'{a}}lffy.
\newblock Waveguide qed with m{\"{o}}ssbauer nuclei.
\newblock {\em Physical Review A}, 109(6):063702, June 2024.

\bibitem{Lohse__2024}
Leon~M. Lohse, Petar Andreji{\'{c}}, Sven Velten, Malte Vassholz, Charlotte
  Neuhaus, Ankita Negi, Anjali Panchwanee, Ilya Sergeev, Adriana P{\'{a}}lffy,
  Tim Salditt, and Ralf R{\"{o}}hlsberger.
\newblock Collective nuclear excitation dynamics in mono-modal x-ray
  waveguides.
\newblock March 2024.

\bibitem{Roehlsberger__2021}
Ralf R{\"{o}}hlsberger and J{\"{o}}rg Evers.
\newblock Quantum optical phenomena in nuclear resonant scattering.
\newblock In Yutaka Yoshida and Guido Langouche, editors, {\em Modern
  M{\"{o}}ssbauer Spectroscopy}, Topics in Applied Physics, pages 105--171.
  Springer Singapore, 2021.

\bibitem{Roehlsberger_S_2010}
R.~R{\"{o}}hlsberger, K.~Schlage, B.~Sahoo, S.~Couet, and R.~R{\"{u}}ffer.
\newblock Collective lamb shift in single-photon superradiance.
\newblock {\em Science}, 328(5983):1248--1251, May 2010.

\bibitem{Heeg_N_2021}
Kilian~P. Heeg, Andreas Kaldun, Cornelius Strohm, Christian Ott, Rajagopalan
  Subramanian, Dominik Lentrodt, Johann Haber, Hans-Christian Wille, Stephan
  Goerttler, Rudolf R{\"{u}}ffer, Christoph~H. Keitel, Ralf R{\"{o}}hlsberger,
  Thomas Pfeifer, and J{\"{o}}rg Evers.
\newblock Coherent x-ray-optical control of nuclear excitons.
\newblock {\em Nature}, 590(7846):401--404, February 2021.

\bibitem{Khairulin_SR_2021}
I.~R. Khairulin, Y.~V. Radeonychev, V.~A. Antonov, and Olga Kocharovskaya.
\newblock Acoustically induced transparency for synchrotron hard x-ray photons.
\newblock {\em Scientific Reports}, 11(1), April 2021.

\bibitem{Bocklage_SA_2021}
Lars Bocklage, Jakob Gollwitzer, Cornelius Strohm, Christian~F. Adolff, Kai
  Schlage, Ilya Sergeev, Olaf Leupold, Hans-Christian Wille, Guido Meier, and
  Ralf R{\"{o}}hlsberger.
\newblock Coherent control of collective nuclear quantum states via transient
  magnons.
\newblock {\em Science Advances}, 7(5), January 2021.

\bibitem{Shvyd’ko_N_2023}
Yuri Shvyd'ko, Ralf R{\"{o}}hlsberger, Olga Kocharovskaya, J{\"{o}}rg Evers,
  Gianluca~Aldo Geloni, Peifan Liu, Deming Shu, Antonino Miceli, Brandon Stone,
  Willi Hippler, Berit Marx-Glowna, Ingo Uschmann, Robert Loetzsch, Olaf
  Leupold, Hans-Christian Wille, Ilya Sergeev, Miriam Gerharz, Xiwen Zhang,
  Christian Grech, Marc Guetg, Vitali Kocharyan, Naresh Kujala, Shan Liu,
  Weilun Qin, Alexey Zozulya, J{\"{o}}rg Hallmann, Ulrike Boesenberg, Wonhyuk
  Jo, Johannes M{\"{o}}ller, Angel Rodriguez-Fernandez, Mohamed Youssef, Anders
  Madsen, and Tomasz Kolodziej.
\newblock Resonant x-ray excitation of the nuclear clock isomer 45sc.
\newblock {\em Nature}, 622(7983):471--475, September 2023.

\bibitem{Velten_SA_2024}
Sven Velten, Lars Bocklage, Xiwen Zhang, Kai Schlage, Anjali Panchwanee,
  Sakshath Sadashivaiah, Ilya Sergeev, Olaf Leupold, Aleksandr~I. Chumakov,
  Olga Kocharovskaya, and Ralf R{\"{o}}hlsberger.
\newblock Nuclear quantum memory for hard x-ray photon wave packets.
\newblock {\em Science Advances}, 10(26), June 2024.

\bibitem{Young_PTotRSoL_1802}
Thomas Young.
\newblock On the theory of light and colours.
\newblock {\em Philosophical Transactions of the Royal Society of London},
  92:12--48, December 1802.

\bibitem{Sturhahn_PRB_2001}
W.~Sturhahn.
\newblock Phase problem in synchrotron m{\"{o}}ssbauer spectroscopy.
\newblock {\em Physical Review B}, 63(9):094105, January 2001.

\bibitem{Sturhahn_ELE_2004}
W.~Sturhahn, C.~L{\textquoteright}abb{\'{e}}, and T.~S. Toellner.
\newblock Exo-interferometric phase determination in nuclear resonant
  scattering.
\newblock {\em Europhysics Letters (EPL)}, 66(4):506--512, May 2004.

\bibitem{Callens_PRB_2005}
R.~Callens, C.~L{\textquoteright}abb{\'{e}}, J.~Meersschaut, I.~Serdons,
  W.~Sturhahn, and T.~S. Toellner.
\newblock Phase determination in nuclear resonant scattering using a velocity
  drive as an interferometer and phase shifter.
\newblock {\em Physical Review B}, 72(8):081402, August 2005.

\bibitem{Heeg_PRL_2015}
K.~P. Heeg, C.~Ott, D.~Schumacher, H.-C. Wille, R.~R{\"{o}}hlsberger,
  T.~Pfeifer, and J.~Evers.
\newblock Interferometric phase detection at x-ray energies via fano resonance
  control.
\newblock {\em Physical Review Letters}, 114(20), May 2015.

\bibitem{Goerttler_PRL_2019}
Stephan Goerttler, Kilian Heeg, Andreas Kaldun, Patrick Reiser, Cornelius
  Strohm, Johann Haber, Christian Ott, Rajagopalan Subramanian, Ralf
  R{\"{o}}hlsberger, J{\"{o}}rg Evers, and Thomas Pfeifer.
\newblock Time-resolved sub-{{\AA}}ngstr{\"{o}}m metrology by temporal phase
  interferometry near x-ray resonances of nuclei.
\newblock {\em Physical Review Letters}, 123(15), October 2019.

\bibitem{Wolff_PRR_2023}
Lukas Wolff and J{\"{o}}rg Evers.
\newblock Unraveling time- and frequency-resolved nuclear resonant scattering
  spectra.
\newblock {\em Physical Review Research}, 5(1):013071, February 2023.

\bibitem{Yuan_NC_2025}
Ziyang Yuan, Hongxia Wang, Zhiwei Li, Tao Wang, Hui Wang, Xinchao Huang,
  Tianjun Li, Ziru Ma, Linfan Zhu, Wei Xu, Yujun Zhang, Yu~Chen, Ryo Masuda,
  Yoshitaka Yoda, Jianmin Yuan, Adriana P{\'{a}}lffy, and Xiangjin Kong.
\newblock Nuclear phase retrieval spectroscopy using resonant x-ray scattering.
\newblock {\em Nature Communications}, 16(1), March 2025.

\bibitem{Negi__2025}
Ankita Negi, Leon~Merten Lohse, Sven Velten, Ilya Sergeev, Olaf Leupold,
  Sakshath Sadashivaiah, Dimitrios Bessas, Aleksandr Chumakhov, Christina
  Brandt, Lars Bocklage, Guido Meier, and Ralf Röhlsberger.
\newblock Energy time ptychography for one-dimensional phase retrieval, 2025.

\bibitem{Krueger_JoSR_2012}
S.~P. Kr{\"{u}}ger, H.~Neubauer, M.~Bartels, S.~Kalbfleisch, K.~Giewekemeyer,
  P.~J. Wilbrandt, M.~Sprung, and T.~Salditt.
\newblock Sub-10nm beam confinement by x-ray waveguides: design, fabrication
  and characterization of optical properties.
\newblock {\em Journal of Synchrotron Radiation}, 19(2):227--236, January 2012.

\bibitem{Fuhse_APL_2004}
Christian Fuhse, Ansgar Jarre, Christoph Ollinger, Jens Seeger, Tim Salditt,
  and Remi Tucoulou.
\newblock Front-coupling of a prefocused x-ray beam into a monomodal planar
  waveguide.
\newblock {\em Applied Physics Letters}, 85(11):1907--1909, September 2004.

\bibitem{Rueffer_HI_1996}
Rudolf R{\"{u}}ffer and Aleksandr~I. Chumakov.
\newblock Nuclear resonance beamline at esrf.
\newblock {\em Hyperfine Interactions}, 97–98(1):589--604, December 1996.

\bibitem{Potapkin_JoSR_2012}
Vasily Potapkin, Aleksandr~I. Chumakov, Gennadii~V. Smirnov, Jean-Philippe
  Celse, Rudolf R{\"{u}}ffer, Catherine McCammon, and Leonid Dubrovinsky.
\newblock The 57fe synchrotron m{\"{o}}ssbauer source at the {ESRF}.
\newblock {\em Journal of Synchrotron Radiation}, 19(4):559--569, May 2012.

\bibitem{Kupenko_HPR_2024}
I.~Kupenko, X.~Li, S.~C. M{\"{u}}ller, D.~Bessas, S.~Yaroslavtsev, G.~Aprilis,
  A.~I. Chumakov, J.-P. Celse, and R.~R{\"{u}}ffer.
\newblock Nuclear resonance techniques for high-pressure research: example of
  the id18 beamline of the european synchrotron radiation facility.
\newblock {\em High Pressure Research}, 44(3):310--336, July 2024.

\bibitem{Smirnov_PRA_2011}
G.~V. Smirnov, A.~I. Chumakov, V.~B. Potapkin, R.~R{\"{u}}ffer, and S.~L.
  Popov.
\newblock Multispace quantum interference in a {57Fe} synchrotron
  m{\"{o}}ssbauer source.
\newblock {\em Physical Review A}, 84(5):053851, November 2011.

\bibitem{Turecek_JoI_2016}
D.~Turecek, J.~Jakubek, and P.~Soukup.
\newblock {USB} 3.0 readout and time-walk correction method for timepix3
  detector.
\newblock {\em Journal of Instrumentation}, 11(12):C12065--C12065, December
  2016.

\bibitem{Gelman__2004}
Andrew Gelman, John~B. Carlin, Hal~S. Stern, and Donald~B. Rubin, editors.
\newblock {\em Bayesian data analysis}.
\newblock Texts in statistical science. Chapman \& Hall/CRC, Boca Raton, Fla.
  [u.a.], 2. ed. edition, 2004.
\newblock Literaturverz. S. 611 - 646.

\bibitem{ForemanMackey_PotASotP_2013}
Daniel Foreman-Mackey, David~W. Hogg, Dustin Lang, and Jonathan Goodman.
\newblock emcee: The {MCMC} hammer.
\newblock {\em Publications of the Astronomical Society of the Pacific},
  125(925):306--312, March 2013.

\bibitem{Lohse_OE_2024}
Leon~M. Lohse and Petar Andreji{\'{c}}.
\newblock Nano-optical theory of planar x-ray waveguides.
\newblock {\em Optics Express}, 32(6):9518, March 2024.

\bibitem{Poikela_JoI_2014}
T.~Poikela, J.~Plosila, T.~Westerlund, M.~Campbell, M.~De Gaspari, X.~Llopart,
  V.~Gromov, R.~Kluit, M.~van Beuzekom, F.~Zappon, V.~Zivkovic, C.~Brezina,
  K.~Desch, Y.~Fu, and A.~Kruth.
\newblock Timepix3: a 65k channel hybrid pixel readout chip with simultaneous
  toa/tot and sparse readout.
\newblock {\em Journal of Instrumentation}, 9(05):C05013--C05013, May 2014.

\bibitem{Saleh__2019}
Bahaa E.~A. Saleh and Malvin~Carl Teich.
\newblock {\em Fundamentals of photonics}.
\newblock Wiley, Hoboken, NJ, third edition. edition, 2019.

\bibitem{Born__2019}
Max Born and Emil Wolf.
\newblock {\em Principles of optics}.
\newblock Cambridge University Press, Cambridge, seventh anniversary edition,
  60th anniversary of first edition, 20th anniversary of seventh edition
  edition, 2019.

\end{thebibliography}

\end{document}